\documentclass[twocolumn,epjc3]{svjour3}  
\journalname{Eur. Phys. J. C}
\usepackage{amsmath,amssymb,graphics,graphicx,multirow,makecell}
\usepackage[style=nature, backend=biber]{biblatex}
\usepackage{subfig}

\let\oldhat\hat
\renewcommand{\vec}[1]{\mathbf{#1}}
\renewcommand{\hat}[1]{\oldhat{\mathbf{#1}}}

\newcommand{\onbb}{$0\nu\beta\beta$}

\newcommand{\gerda}{\textsc{Gerda}}
\newcommand{\legend}{\textsc{Legend}}
\newcommand{\majorana}{\textsc{Majorana Demonstrator}}
\newcommand{\mjd}{\textsc{MJD}}

\usepackage{lineno}

\begin{document}

\title{Charge-carrier collective motion in germanium detectors for $\beta\beta$-decay searches
}


\author{Tommaso Comellato\thanksref{addr1,e1}
        \and
        Matteo Agostini\thanksref{addr1, addr2, e2}
        \and
        Stefan Schönert\thanksref{addr1,e3}
}

\thankstext{e1}{e-mail: tommaso.comellato@tum.de}
\thankstext{e2}{e-mail: matteo.agostini@ph.tum.de}
\thankstext{e3}{e-mail: schoenert@ph.tum.de}



\institute{
        Physik Department E15, Technische Universität München, James-Franck-Straße 1, 85748, Garching, Germany \label{addr1}
        \and
        \emph{Present address:} Department of Physics and Astronomy, University College London, Gower Street, London WC1E 6BT, UK \label{addr2}
}

\date{Received: 3 August 2020 / Accepted: 17 January 2021}

\maketitle

\begin{abstract}

The time analysis of the signal induced by the drift of charge carriers in high purity germanium detectors provides information on the event topology. Millions of charge carriers are produced in a typical event. Their initial distribution, stochastic diffusion and Coulomb self-repulsion affect the time structure of the signal. We present a comprehensive study of these effects and evaluate their impact on the event discrimination capabilities for the three geometries which will be used in the \legend~experiment for neutrinoless double-beta decay.

\end{abstract}

\section{Introduction}
\label{intro}
Since the invention of transistors in 1948 \cite{Bardeen1948}, germanium has been used in a broad variety of applications, ranging from gamma-ray detection \cite{HPGeDet} to fiber optics \cite{friebeleGriscom:defect, ballato:glassclad} to search for dark matter \cite{MJD:DM2017, cogent:APS, CDMS2020}. The state-of-the-art technology allows the production of detector blanks with lengths and diameters of 8-9 cm using the Czochralski method. With a level of impurities of the order of $10^{10}$ atoms/cm$^3$, such crystals can be converted into High Purity Germanium (HPGe) detectors. A HPGe detector is a semiconductor device. Two electrodes on the crystal surface are used to apply a bias voltage and extend the semiconductor junction throughout the full detector volume. 
When a gamma-ray or charged particle interacts within the detector it creates a large number of charge carriers, i.e. electrons and holes. Charge carriers of the same sign drift together towards the electrodes as cluster, following the electric field lines. Their motion induces a signal on
the electrodes that is typically read-out by a charge sensitive amplifier. Similar to a time projection chamber, the analysis of the time structure of the read-out signal contains information on the topology of the event, i.e. on the number and location of the energy depositions. 

An important field of applications for germanium detectors is the search for neutrinoless $\beta\beta$ decay (\onbb), a nuclear transition predicted by many extensions of the Standard Model of particle physics in which two neutrons decay simultaneously into two protons and two electrons. For this search, detectors are fabricated from germanium material isotopically enriched to $\sim$90\% in the candidate double-beta decaying isotope $^{76}$Ge. Thus, the decay occurs inside the detector and the electrons are absorbed within $\mathcal{O}(mm)$, producing a point-like energy deposition. For \onbb~experiments it is hence of primary interest to discriminate single-site energy depositions (typical of the sought-after signal) from multiple-site energy depositions (typical of background events induced by multi-Compton scattering), as well as surface events (which, for geometrical reasons, are more likely to be external $\alpha$ or $\beta$ particles).

The time development of the signal depends on the geometry of the detector, its electrode scheme, and its impurity concentration. Thus, an accurate modeling of the signal formation and evolution is an essential ingredient to design the detector and enhance the accuracy of the topology reconstruction and event discrimination. As an example, simulations have been extensively used in gamma-spectroscopy, such as modeling the segmented detectors of AGATA and GRETA~\cite{agatagretinareview, agata:pulseshape}, while in \onbb~experiments they led to Broad Energy Germanium (BEGe) and P-type Point Contact (PPC) detectors \cite{DusanBEGe, MJDExperienceWGermanium}.
In the effort to increase the detector mass, new geometries such as the Inverted Coaxial (IC) \cite{RadfordCooperIC} have recently drawn increasing attention. In this new type of detectors, the time needed to collect electrons and holes is much longer than in the aforementioned geometries. 

In this article we investigate the collective effects in a cluster of charge carriers and their impact on the signal formation in the detector geometries of interest for \onbb~searches. Section \ref{sec:BasicModeling} summarizes the charge-carrier collection and signal formation for the detector geometries under consideration. Section \ref{sec:CollectiveEffects} describes collective effects in charge-carriers' clusters, which include self-repulsion, thermal diffusion and velocity dispersion. Section \ref{sec:impact0nbb} discusses the impact of such effects on the signal and background discrimination in \onbb~searches and Section \ref{sec:conclusion} finally discusses the results and puts them in the context of the future \legend~experiment. 

We performed comprehensive simulations of germanium detectors and validated them against the data acquired with a custom designed IC detector produced in collaboration with Baltic Scientific Instruments (BSI) and Helmholtz Research Center (Rossendorf). Its geometry is the one used as reference for this paper. Our work builds on the results of~\cite{Mertens}, which reports the first observation of such effects in PPC detectors and discusses how to accurately model them. Our simulations have been carried out with the \textsc{Mage}~\cite{MaGe} software framework based on \textsc{Geant-4}~\cite{geant4}, and a modified version of the SigGen software package~\cite{Radware} which already included the modeling of the collective effects and was used in~\cite{Mertens}. More details on simulations are given in \ref{app:sim}.



\section{Charge-carrier collection and signal formation in germanium detectors}
\label{sec:BasicModeling}
\begin{figure*}[ht]
  \includegraphics[width=\textwidth]{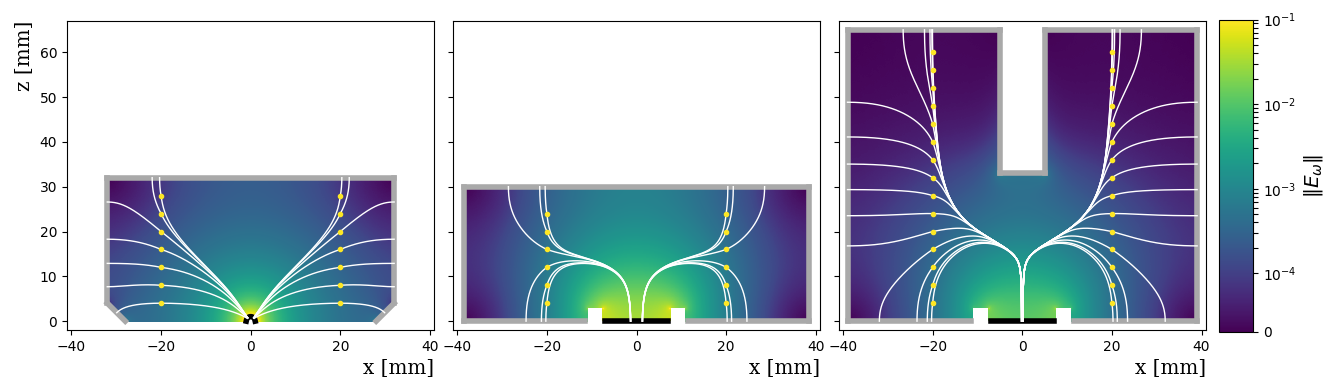}
\caption{Weighting field $E_\omega$ for a cross section of the three geometries used in current and future \onbb~experiments: (from left) PPC, BEGe and inverted coaxial. The thick black and gray lines are the p$^+$ and n$^+$ electrode, respectively. The yellow points are locations of an energy deposition, the white trajectories connecting them to the p$^+$ electrode are the drift paths of holes and those connecting them to the n$^+$ electrode are the drift paths of electrons.}
\label{fig:Wfield}       
\end{figure*}

When gamma-rays or charged particles interact within the germanium detector they release energy. About $10^6$ electron-hole pairs are created for each MeV released in the active detector volume. Once produced, the two kinds of carriers drift as two clusters in opposite directions following the electric field lines until they reach the electrodes. The signal induced by the motion of these charges can be to a first approximation modeled by the Shockley–Ramo theorem \cite{Shockley, Ramo}. The theorem states that the instantaneous current $I(t)$ induced at a given electrode by a drifting cluster of charge q is given by
\begin{equation}
    I(t) = q \, \vec{v} (\vec{x}(t)) \cdot \vec{E_\omega} (\vec{x}(t))
    \label{eq:ShockleyRamo}
\end{equation}
where $\vec{v} (\vec{x}(t))$ is the instantaneous drift velocity and $\vec{E_\omega}(\vec{x}(t))$ is the weighting field at position $\vec{x}(t)$. The weighting field is defined as the electric field created by the considered electrode set at 1 V, all other electrodes grounded and all charges inside the device removed. Thus, the signal induced at the electrode is the product of the instantaneous drift velocity and the projection of the weighting field in the direction of motion, weighted by the deposited charge. 

Often events induced by gamma-rays result in multiple energy depositions well separated compared to the dimension of the charge clusters. In this case, each cluster drifts independently of the others and the resulting signal is the superposition of the individual signals, each of them weighted by the charge in each cluster.

Three illustrative HPGe detector geometries are analyzed in this article. These are the geometries used by the current and future \onbb~experiments: \gerda~\cite{Gerda:science}, \majorana~(\mjd) \cite{MJD:PRL2019}, \legend~\cite{Legend:Medex2017}. All of them are p-type detectors, with a Lithium-diffused n$^+$ electrode and a B-implanted p$^+$ electrode. The three detector types are shown in Fig.~\ref{fig:Wfield} along with the resulting weighting field and illustrative trajectories.

The PPC detectors have a cylindrical shape and have masses up to 1 kg. Their geometry is characterized by a small ($\sim$2 mm diameter) p$^+$ electrode on one of the flat surfaces, while the rest of that flat surface is passivated. The remaining surface of the detector is covered by the n$^+$ electrode. Electrons are collected on the n$^+$ electrode that is kept at a few kV operational voltage, while holes on the p$^+$ electrode, that is grounded and used to read-out the signal. This geometry creates a weighting field that increases rapidly in the immediate vicinity of the p$^+$ electrode. This results in a characteristic peak-like structure in the current signal when the hole clusters approach the p$^+$ electrode. 

Compared to PPC detectors, the BEGe detectors are shorter but have a larger radius. The major difference between the two geometries is the structure of the electrodes: the p$^+$ electrode is larger for BEGe (up to $\sim$15 mm diameter) and surrounded by a passivated groove with typical depths of $\sim$3 mm. The BEGe detectors’ n$^+$ electrode extends down to the groove, wrapping around the crystal on all surfaces. This structure has a strong impact on the trajectories of the carriers, as it creates a \emph{funnel} effect \cite{Agostini_SignalModeling}: holes are pushed towards the center of the detector and then move to the p$^+$ electrode along a fixed path that is independent by their starting point  (see central plot of Fig.~\ref{fig:Wfield}). Since that is the volume in which the weighting field is highest, according to Eq.~\ref{eq:ShockleyRamo}, the majority of the induced signals in a BEGe detector share the same maximum value of the current $I(t)$.

The inverted coaxial detector has the same electrode structure as a BEGe, though it is about twice as long. In order to keep a high electric field throughout the whole volume, a hole is drilled on the opposite side of the p$^+$ electrode and constitutes part of the n$^+$ contact. It normally extends down to within 25-35 mm from the p$^+$ electrode. With the wrap-around n$^+$ electrode, the funneling is preserved and the trajectories converge in the region of high weighting field (see Fig.~\ref{fig:Wfield}).

\section{Charge-carrier collective effects}
\label{sec:CollectiveEffects}

The modeling of the signal formation presented in the previous section does not account for the cluster spatial extension that is $\mathcal{O}(mm)$ for a MeV energy deposition. It can be extended to account for the non-null dimensions of the cluster. If we define $\vec{r}(t)$ as the distance of every charge in the cluster from the center of the distribution, the instantaneous signal induced at the electrode will be the integral of equation \ref{eq:ShockleyRamo} over the spatial charge distribution $Q(\vec{r}(t))$ of the cluster:
\begin{equation}
    \widetilde{I}(t) = \int d\vec{r} \, Q(\vec{r}(t)) I(t).
    \label{eq:extendedSR}
\end{equation}
If the electric field varies on scales similar to the cluster size, charges at the opposite side of the cluster will experience different forces (accelerations), leading to a  deformation of the cluster during its drift towards the electrodes.
Moreover, the stochastic diffusion and self-interaction of the charge carriers will progressively increase the size of the cluster during its motion. The diffusion consists of a random thermal motion of the carriers while the self-interaction is the result of the Coulomb force. In this work, such processes are treated as collective effects~\cite{Radware}. That allows an analytical treatment and  keeps the computational requirements to an affordable level. We compared this approximated collective description with a full multi-body simulation\footnote{We simulated the individual motion of 10000 charges in the field generated by the detector and the instantaneous configuration of the other charges.} and found that it does not introduce noticeable inaccuracies. 

In our collective treatment, we consider the effects of mutual repulsion and diffusion separately from those of acceleration, because the formers act in all directions, while the latter breaks the spherical symmetry and acts exclusively in the direction of motion.
 
The dynamics of drifting charges in the presence of mutual repulsion and diffusion can be treated assuming spherical symmetry,  and is described by the continuity equation \cite{gatti}:
\begin{equation}
    \frac{\partial^2 Q}{\partial r^2} - \frac{2}{r}\frac{\partial Q}{\partial r} - 
    \frac{1}{D} \frac{\partial Q}{\partial t} - Q \frac{\partial Q}{\partial r}\frac{1}{V_T} \frac{1}{4\pi\epsilon r^2} = 0
    \label{eq:contEq}
\end{equation}
where $D$ is the diffusion coefficient, $\epsilon$ the permittivity in germanium and $V_T$ the thermal voltage $V_T = k_B T/q$ with $q$ being the elementary charge. The general solution of Eq. \ref{eq:contEq} when the Coulomb repulsion term is neglected describes a gaussian profile for the charge cluster, whose width is 
\begin{equation}
    \sigma_D = \sqrt{2Dt}.
    \label{eq:diff}
\end{equation} 
When charges drift in an electric field, the diffusion coefficient $D$ has a longitudinal and transverse component. Both are calculated in SigGen~\cite{Radware} in the respective direction, but only the longitudinal is the responsible for the deformation of the signal. As reported in~\cite{jacoboni}, this component is lower as the electric field strength increases. This implies that, with a sufficiently high impurity concentration, the effect of diffusion can be strongly limited (as stated also in~\cite{Mertens}).

Neglecting the first two terms of Eq. \ref{eq:contEq} and considering only the Coulomb self-repulsion, we obtain a solution in which the charge distribution behaves like an expanding sphere of radius $\sigma_R$:
\begin{equation}
    \sigma_R = \sqrt[3]{\frac{3 \mu q}{4\pi \epsilon} N t}
    \label{eq:rep}
\end{equation} 
where N is the number of charge carriers in the distribution and $\mu$ is the mobility of the carrier, which is related to the diffusion coefficient by the Einstein equation $D=\mu k_BT/q$. 

Both Eq. \ref{eq:diff} and \ref{eq:rep} describe a distribution which gets monotonously broader with time, with the difference that Eq. \ref{eq:diff} is completely determined by the detector properties, while Eq. \ref{eq:rep} depends on the deposited energy. The drifting in the electric field of the detector, on the other hand, enlarges or decreases the size of the cluster, according to whether it experiences accelerations or decelerations. The modeling of such effect is obtained from basic kinematics, and can be easily calculated for each time-step $t_i$ as:
\begin{equation}
	\sigma_A(t_{i+1})  = \sigma_A(t_{i}) \cdot \frac{v(t_{i+1})}{v(t_i)}
\end{equation}
It is clear that in the direction of motion there is a strong interplay between the three described effects, which can give rise to non-linear effects on the cluster size.
\begin{figure*}[]
  \includegraphics[width=\textwidth]{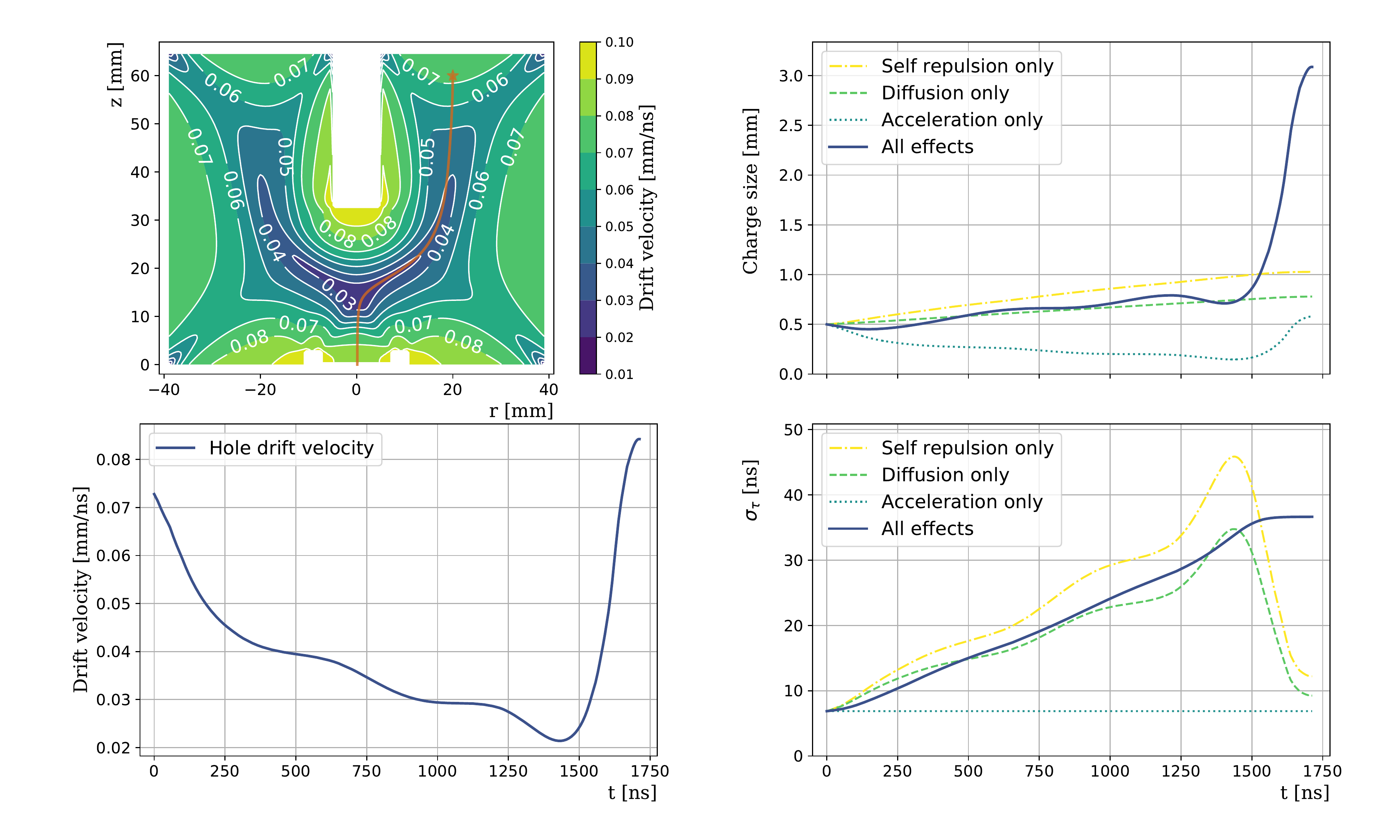}
\caption{Breakdown of the collective effects on a charge cluster. The top-left plot shows the drift velocity field of an IC detector with superimposed in brown the drift path of the holes' cluster for an interaction location marked by the star. The cluster's drift velocity along the path is shown in the bottom-left plot. The evolution of the cluster's size and $\sigma_\tau$ is displayed in the top-right and bottom-right plot, respectively. The initial size of the cluster is 0.5 mm, the average for energy depositions of 1.6 MeV.}
\label{fig:breakdown}       
\end{figure*}

Fig. \ref{fig:breakdown} displays the contribution of the mentioned processes to the charge cluster deformation\footnote{The initial cluster size is given here in Full Width Half Maximum, and it has been determined as a function of energy through Monte Carlo simulation. See details in~\ref{app:sim}.}. The top-left plot shows the drift velocity field on an IC detector cross section, where superimposed in brown is the trajectory of holes for an energy deposition on the position marked with the star. As holes travel through the detector, they experience accelerations (decelerations) according to the electric field, stretching (shrinking) the cluster size in the direction of motion as shown in the top-right panel (light blue curve). In the same plot, the broadening effect due to the described Coulomb and diffusion processes are shown with the yellow and green curves, respectively: as described by equations \ref{eq:diff} and \ref{eq:rep}, their effect is a monotonic enlargement of the cluster size. Finally, the dark blue curve shows the evolution of the cluster dimensions, when all effects act simultaneously. As anticipated, the total size is not just the simple sum of the three contributions, as they are not independent: an enlargement of the cluster size, for instance due to Coulomb or diffusion effects, emphasizes the difference in the drift velocity field of charges at the edge of the distribution, thus amplifying the effect of acceleration. This amplification effect has been tested with our full multi-body simulation mentioned above. In our multi-body simulation, we calculated the motion of every single charge induced by the field created by the detector, superimposed to the field created by the other charges in the cluster. That approach confirmed the evolution of the cluster size as modeled by the collective description presented above. In particular, it reproduces the amplification effect of acceleration and mutual repulsion, thus further confirming the modeling in SigGen.

The impact of the different collective effects on the signal formation can be characterized through the time spread of the cluster, which we define in the following as $\sigma_\tau(t)$. 
The evolution in time of such parameter is displayed in the bottom right plot of Fig.~\ref{fig:breakdown}. The light blue curve shows that $\sigma_\tau$ is constant if only acceleration effects are considered. As other effects are switched on, their interplay gives a total time spread which can be up to a factor 5 larger than the initial value. 

\begin{figure*}
  \includegraphics[width=\textwidth]{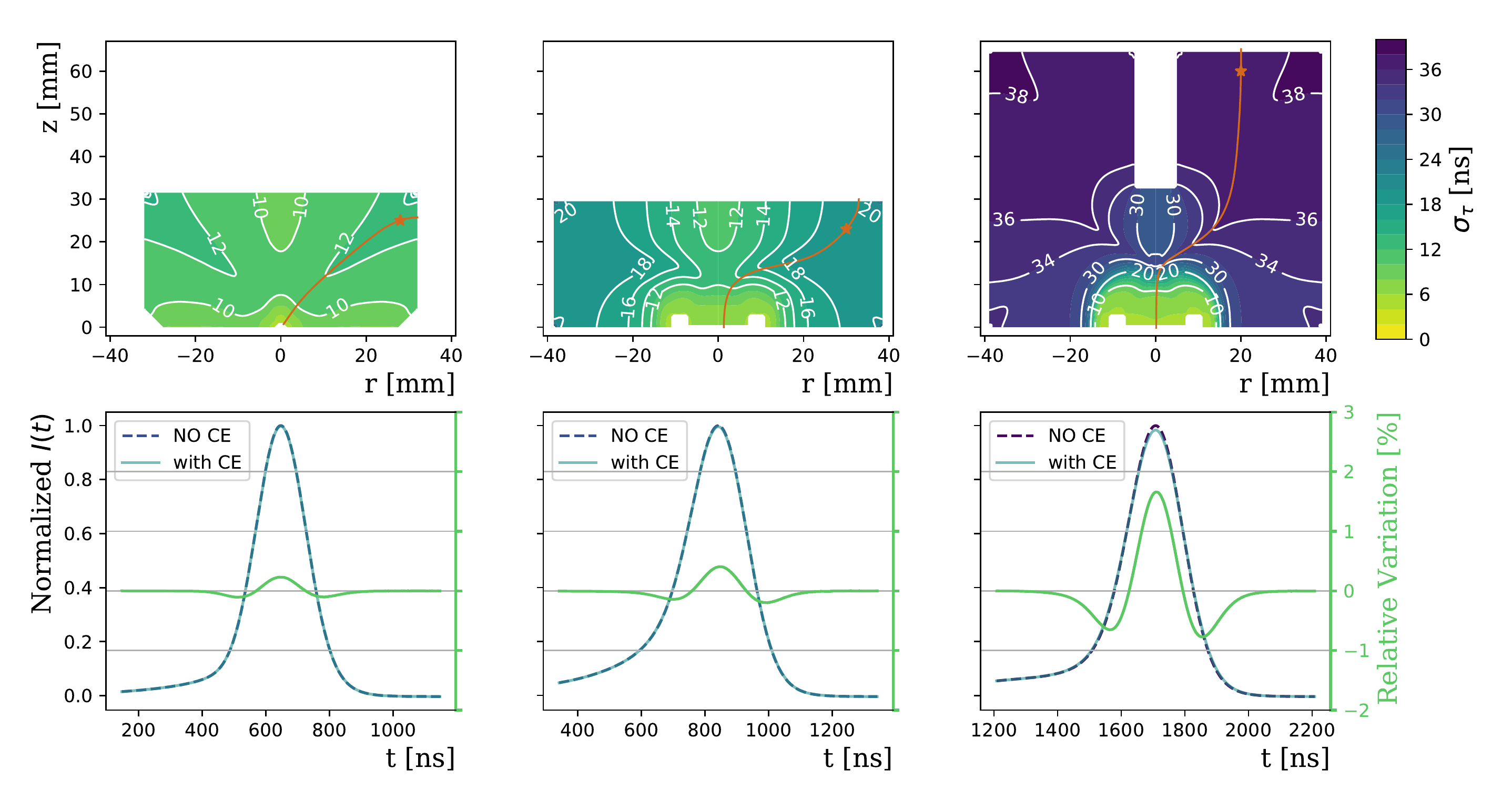}
\caption{Top: values of the $\sigma_\tau$ parameter as a function of the interaction position, for the three geometries considered. Bottom: simulated signals for the interactions and drift paths indicated by the brown point and curve, with and without Collective Effects (CE). Higher values of $\sigma_\tau$, as in inverted coaxial detectors, imply lower values of the current $I(t)$.}
\label{fig:DeltaTau&Pulse}       
\end{figure*}

The enlargement of the cluster size through the parameter $\sigma_\tau$ as a function of the interaction position is shown in Fig. \ref{fig:DeltaTau&Pulse} (top), separately for the three considered geometries. For PPC detectors, the maximum enlargement is for interactions in the corners, where $\sigma_\tau$ reaches about 15 ns. The corners are the part of the detector from which the hole drift path is the longest. For BEGe detectors the maximum is slightly larger, up to 20 ns for radii larger than 30 mm. For inverted coaxial detectors the effect is much stronger, up to a factor 2 and it affects more than half of the detector volume. The impact on the signal shape is shown in the bottom row of Fig. \ref{fig:DeltaTau&Pulse}, where signals are shown with (light blue) and without (dark blue) the deformation caused by collective effects. The difference between the two cases is less than 0.5$\%$ of the signal amplitude in BEGe and PPC detectors (see green curve), but it is larger for inverted coaxials, where the maximum of the current signal is lowered by $\sim$2 \% when group effects are switched on.

The collective effects described in this section are expected for all detector geometries. Their impact on the signal shape, however, will depend on the geometry and the impurity profile. 
In the second part of this paper, we will evaluate such impact on advanced event reconstruction techniques such as those for \onbb~experiments.

\section{Event discrimination in \onbb~experiments}
\label{sec:impact0nbb}

\onbb~experiments using HPGe detectors rely heavily on the analysis of the time structure of the signal in order to reconstruct the topology of the energy deposition and thus discriminate between \onbb~and background. This kind of analysis is commonly referred to as Pulse Shape Analysis (PSA). \onbb~events are characterized by a single energy deposition while background can be generated by gamma-rays scattering multiple times within the detector, or $\alpha$ and $\beta$ particles depositing energy next to the detector surface\footnote{These surface events generate peculiar pulse-shapes, the recognition of which is beyond the scope of this work.}.
PSA techniques are based on the recognition of a few specific features of the signal time evolution which allows for a discrimination between signal- and background-like events.
The effects discussed in the previous section have the net result of blurring these features and, consequently, of worsening the performance of any PSA technique. In this section we evaluate their impact on a particular PSA technique that is the standard in the field: the so called $A/E$ method~\cite{DusanBEGe}. 

The $A/E$ technique is based on a single parameter that is the maximum value of the current signal ($A$), normalized by the total deposited energy ($E$) (or $q$ in Eq. \ref{eq:ShockleyRamo}). 
In case of a single energy deposition, the signal has a single peak structure with amplitude $A$, which corresponds to the moment when the holes' cluster passes through the region of maximum weighting field.

If the energy is deposited in multiple locations, multiple clusters are simultaneously created and the total signal is the superposition of the signal induced by the motion of each of them.  Different clusters will reach the region of maximum weighting field at different times, creating a multiple peak structure.
Since the amplitude of each peak is proportional to the total charge in the cluster generating it, events with multiple energy depositions $E_i\propto q_i$ will have a lower $A/E$ value compared to single-site events in which all energy is concentrated in a single cluster $E\propto \sum_i q_i$. When normalized to the total charge $q$, the signal from a multiple energy deposition gives lower $A/E$ values compared to a single energy deposition. More details are given in \ref{app:A/E}. 

The $A/E$ parameter is independent of the interaction position and its discrimination efficiency is constant throughout the whole detector volume. This is due to the fact that the holes approach the region of maximum weighting field along the same trajectory\footnote{This is true for BEGe and IC detectors. The funneling effect is not present in the PPCs, because for that geometry the weighting field at the p$^+$ electrode is spherical, hence the signal does not depend on the angle from which the holes arrive.}, independent of the original location where the cluster was created. Without considering the collective effects, the $A/E$ parameter is expected to have the same value for clusters with a given energy generated in most of the detector volume. The only exception is for interactions nearby the read-out electrode, for which the $A/E$ parameter is larger than usual because of extra contribution of the electrons' cluster that now moves in a region of strong electric and weighting field and its contribution on the signal shape is not negligible as in the rest of the detector. The uniformity of the $A/E$ parameter in the detector volume has been studied in detail in~\cite{Agostini_SignalModeling}.
Collective effects depend on the interaction position -- as shown by the $\sigma_\tau$ parameter in Fig. \ref{fig:DeltaTau&Pulse} -- and this creates an $A/E$ dependence from the interaction position.

\begin{figure*}
  \includegraphics[width=\textwidth]{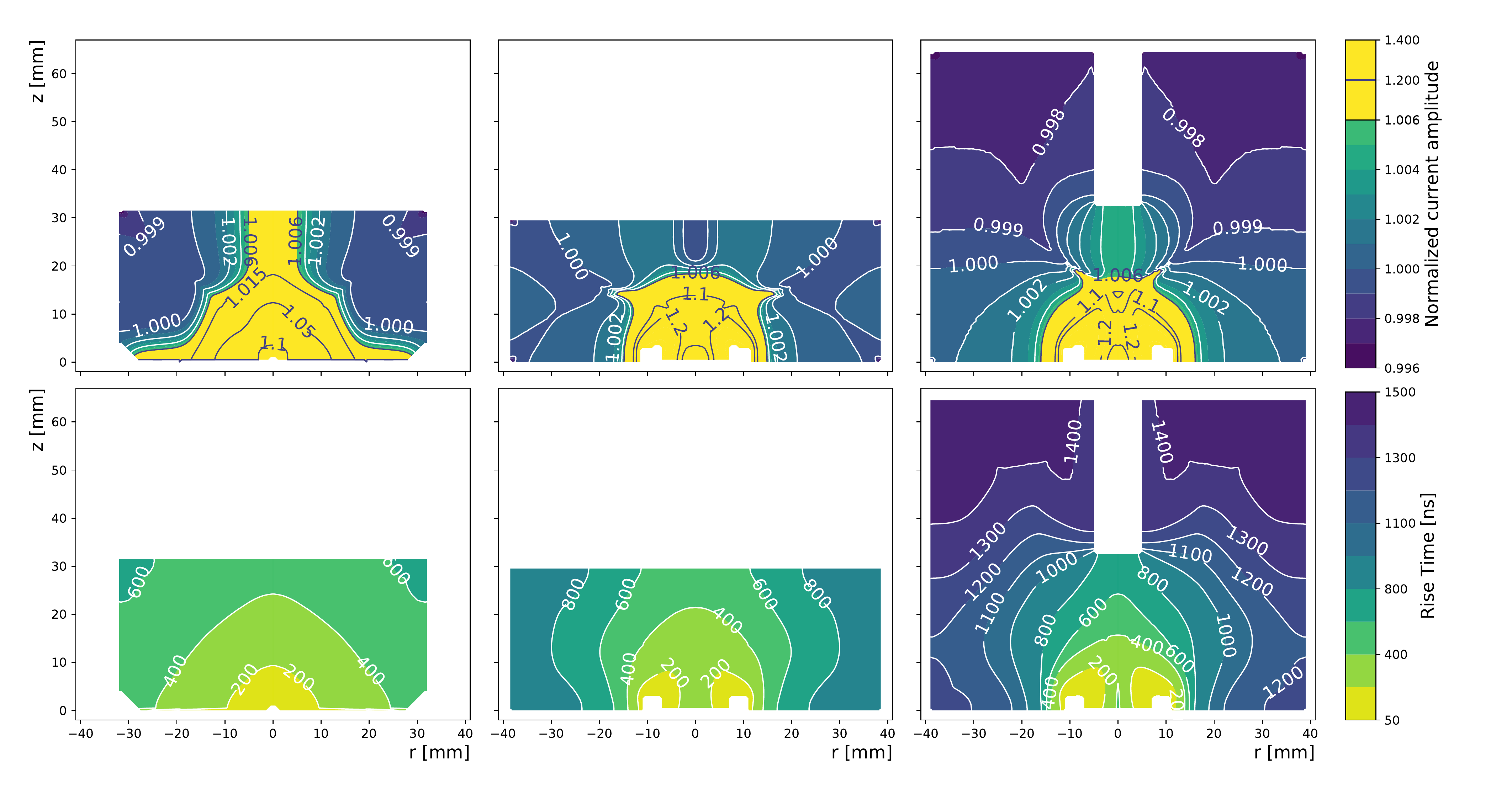}
\caption{$A/E$ (top) and rise time (bottom) values for the three analyzed geometries. In PPC and BEGe detectors rise times range up to 600-800 ns, while for inverted coaxials they can be twice as big, and saturate for high z-positions, where the threshold at 0.5\% is no longer a good approximation of the beginning of charge collection. A correlation between $A/E$ and rise time is visible for the inverted coaxial detector. }
\label{fig:HPGe_ART}
\end{figure*}

Fig. \ref{fig:HPGe_ART} shows the value of the $A/E$ parameter for mono-energetic energy depositions simulated throughout the whole detector volume considering the collective effects described in Sec.~\ref{sec:CollectiveEffects}. The $A/E$ value varies by a few percent between the corners and the center of the detector in the BEGe and PPC geometry. As already mentioned, the value is significantly amplified only in about 3\% of the detector volume around the p+ electrode. For inverted coaxial detectors, while the bottom half of the volume exhibits features similar to the BEGe geometry, the upper part shows a consistent 0.3\% reduction of the $A/E$ value. This reduction of the $A/E$ has been experimentally confirmed by studying the response of our prototype inverted coaxial detector against low-energetic gamma-rays used to create well-localized energy depositions on different parts of the detector surface.  

Maximizing the detector volume is of primary importance for \onbb~experiments. However, the larger the collection path, the stronger the impact of these collective effects will be. In the following we evaluate the event-reconstruction performance of inverted coaxial detectors and discuss possible analysis techniques to correct for these collective effects. To quantify the performance we focus on the acceptance of \onbb-like events and of typical backgrounds of the experiments.

The event discrimination based on the $A/E$ parameter is calibrated using the Double Escape Peak (DEP) events from $^{208}$Tl as a proxy for \onbb~events, as they both consist in a single energy deposition (for more details on the calibration of the analysis, we refer to \ref{app:A/E}). The $A/E$ distribution of DEP events is used to set a cut value which keeps 90\% of their total number. This value cannot be directly translated to \onbb~acceptance, for two reasons: the first is that DEP and \onbb~events have a slight different topology\footnote{DEP events consist in an electron and positron sharing 1.6 MeV, while \onbb~events produce two electrons sharing 2 MeV. This changes the initial cluster size, as well as the Bremsstrahlung probability.}. The second, DEP events are concentrated on corners, \onbb s are homogeneously distributed. 

In order to estimate the \onbb~acceptance, we performed a Monte Carlo simulation of the energy deposited in 300000 \onbb~and DEP events. The Monte Carlo simulation takes into account all the physical differences between the two classes of events and their spatial distribution within the detector. For each event, the total signal is computed using the modeling described in Sec.~\ref{sec:BasicModeling} and \ref{sec:CollectiveEffects} and analyzed to extract the $A/E$ parameter. From the $A/E$ distribution of DEP events, we set the cut value and applied it to the \onbb~population. This resulted in a final \onbb~acceptance of $(86.1\pm 0.1$(stat))\%, which is compatible with the typical values for BEGe detectors \cite{Gerda:science} (see Tab.~\ref{tab:SP}). Technical details on Monte Carlo and pulse shape simulation, as well as on the signal processing can be found in \ref{app:sim}.

From the Monte Carlo simulation of $^{208}$Tl, we also extracted the $A/E$ distributions of events from $^{208}$Tl Full Energy Peak (FEP), $^{208}$Tl Single Escape Peak (SEP) as well as from the Compton continuum (CC) from $^{208}$Tl and $^{214}$Bi, which constitute background at Q$_{\beta\beta}$. We applied the cut obtained from DEP events to these distributions and obtained the survival fraction of $(5.1\pm0.3$)\% and $(7.4\pm0.1$)\% for SEP and FEP events, respectively (see Tab. \ref{tab:SP}), and $(45.1\pm0.3$)\% and $(20.3\pm0.4$)\% for the Compton continuum at Q$_{\beta\beta}$ from $^{208}$Tl and $^{214}$Bi, respectively. The values, reported in Tab. \ref{tab:SP}, are in agreement with the typical theoretical values for BEGe detectors \cite{Agostini_SignalModeling}. 


\begin{table*}
\begin{tabular}{cclclc|clclcl}
\hline\noalign{\smallskip}
                &   \multicolumn{5}{c}{\textbf{Simulations}}  & \multicolumn{6}{c}{\textbf{Data}} \\ 
                &   \multicolumn{4}{c}{IC} & BEGe\cite{Agostini_SignalModeling, Gerda:science} &  \multicolumn{4}{c}{IC (this work)}   &  \multicolumn{2}{c}{BEGe \cite{Agostini_SignalModeling, DusanBEGe} }  \\
                
Event class     & \multicolumn{2}{c}{\emph{Standard}} &  \multicolumn{2}{c}{\emph{RT corr}} & \emph{Standard} & \multicolumn{2}{c}{\emph{Standard}} &  \multicolumn{2}{c}{\emph{RT corr}}& \multicolumn{2}{c}{\emph{Standard}}  \\

\noalign{\smallskip}\hline\noalign{\smallskip}

$^{208}$Tl DEP  &   90.00 & (8) & 90.08 & (8) & 90  \quad (1) & 90.1 & (8) & 90.1 & (8)  &  90 & (1)\\
$^{208}$Tl SEP  &   5.1   & (3) & 5.8   & (3) & 8 \,\ \quad (1) & 5.0  & (3) & 5.3  & (3)  &  5.5  & (6)  \\
$^{208}$Tl FEP  &   7.4   & (1) & 8.1   & (1) & 12  \quad  (2)   & 7.64 & (5) & 7.92 & (5)  &  7.3  & (4)  \\
CC @Q$_{\beta\beta}$ ($^{208}$Tl)& 45.1 &(3) & 46.7 &(3) & 42 \quad (3) & 32.3 &(2)& 33.1 & (2)  & 34 &(1)\\
CC @Q$_{\beta\beta}$ ($^{214}$Bi)& 20.3 &(4) & 21.8  &(4) & -- &  \multicolumn{2}{c}{--}  &  \multicolumn{2}{c}{--} & 21 & (3)   \\

\noalign{\smallskip}\hline\noalign{\smallskip}

\onbb           &   86.07 & (6) & 85.47 &(6) & 88\quad (2)   &\multicolumn{2}{c}{--} & \multicolumn{2}{c}{--} & \multicolumn{2}{c}{--}   \\

\noalign{\smallskip}\hline
\end{tabular}
\caption{Percentage of events classified as single-site for different event samples and detectors, taken from simulations and experimental data. For inverted coaxial detectors, the results are given both before (\emph{Standard}) and after a correction based on the rise time (\emph{RT corr}).}
\label{tab:SP}       
\end{table*}

\begin{figure*}
\centering
\subfloat[][DEP events from data (filled colored contour) and simulations (gray contour lines) ]{\includegraphics[width=.48\textwidth]{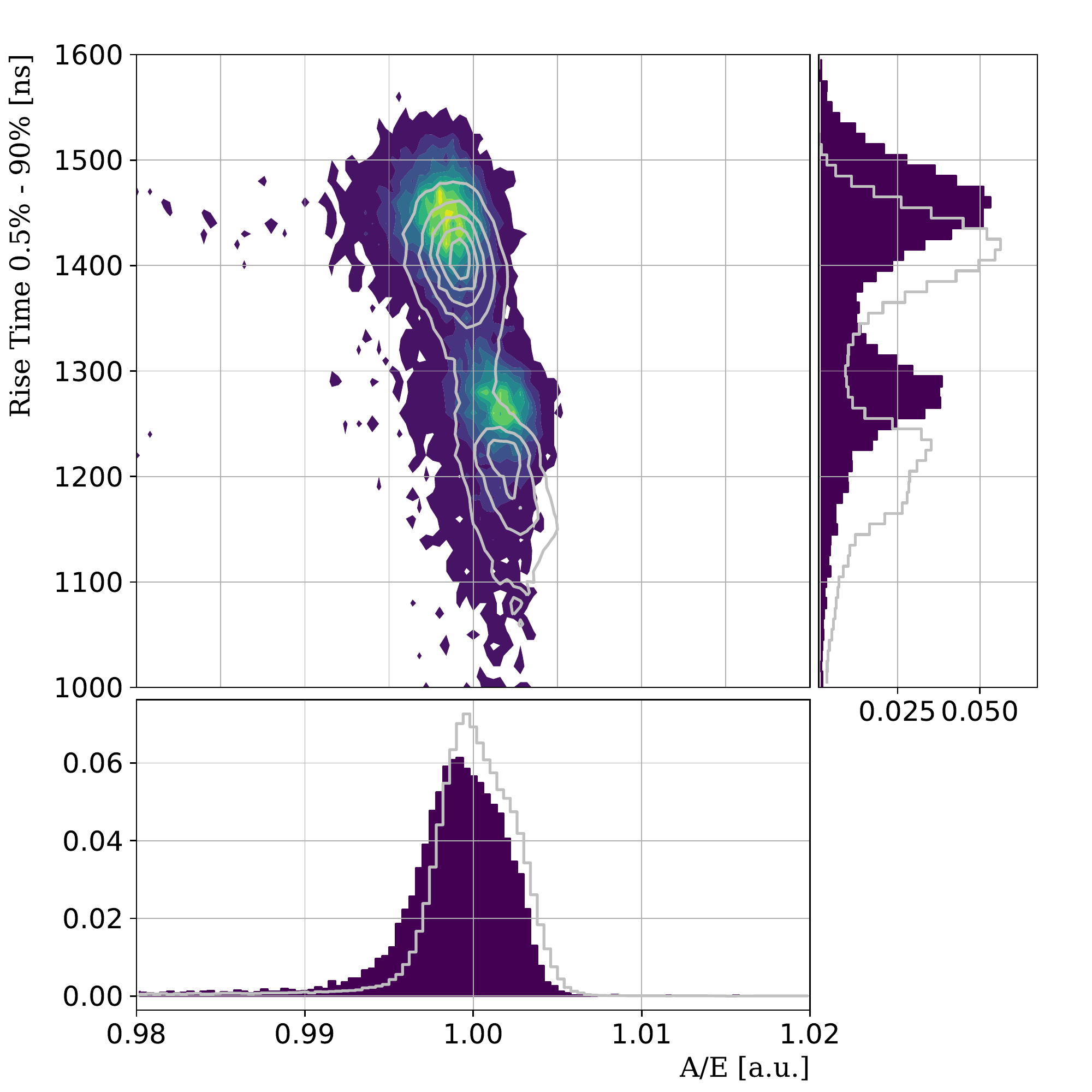}\label{subfig:RTvsA/E_DEP}} \quad
\subfloat[][\onbb~events from simulations ]{\includegraphics[width=.48\textwidth]{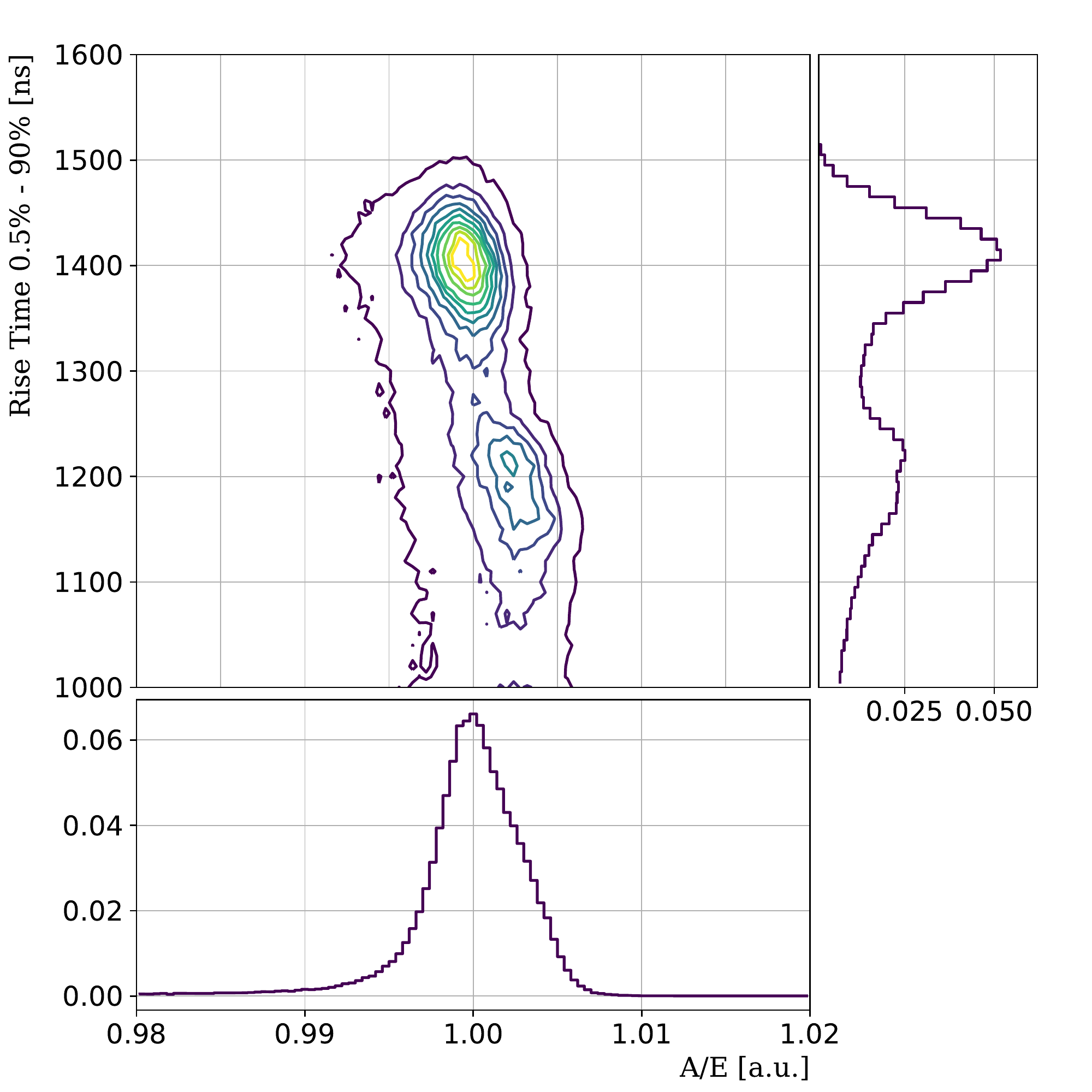}\label{subfig:RTvsA/E_0nbb}} 
\caption{Distribution of the $A/E$ and rise time for (\ref{subfig:RTvsA/E_DEP}) DEP events and (\ref{subfig:RTvsA/E_0nbb}) \onbb~events. The distributions are shown for experimental data (color maps) and simulated data (contour lines).}
\label{fig:RTvsAoE}       
\end{figure*}

As pointed out above, the impact of the collective effects is correlated with the time needed to collect the hole cluster. Following the proposal of \cite{commDavid}, we tested a correction on the $A/E$ parameter based on the reconstructed collection time of the signals, in order to restore the position independence. In this work we reconstruct such a quantity by taking the time between two arbitrary thresholds on the signal, i.e. what is called the rise time\footnote{Normally, the thresholds are set on the signal which is experimentally accessible, which means the output of the charge sensitive pre-amplifier. That is the charge signal $V(t)$, which is the integral of the current signal $I(t)$.}. Noise conditions can prevent accurate determination of the start time for thresholds below 0.5\% at the energies of interest for \onbb~search. Hence, for this work we refer to rise time as the time between 0.5\% and 90\% of signal development\footnote{Other techniques, based on the convolution of the signal function with a well tuned impulse response function, could lead to the identification of lower thresholds, such as 0.1\% of the signal amplitude.}. 
A map of the mean rise time as a function of the interaction position within the detector is shown in Fig. \ref{fig:HPGe_ART} for the three geometries considered. These rise time and $A/E$ values are  correlated in the inverted coaxial geometry. This is shown explicitly in Fig. \ref{fig:RTvsAoE} for DEP (\ref{subfig:RTvsA/E_DEP}) and \onbb~(\ref{subfig:RTvsA/E_0nbb}) events. Both plots suggest that a linear correlation could be used to align the $A/E$ values in the bottom and top part of the detector volume.

This double peak structure has been first reported in \cite{DomulaHult, comellatoPisa}. Its origin is connected by our work to the collective effects and the spatial distribution of DEP events within the detector. Indeed, the configuration of the inverted coaxial detector creates a region on the top and one on the bottom part of the detector in which rise time and $A/E$ saturate to a limit value, which gets more represented than the others. This effect is even more pronounced for DEP events, which are more likely to occur on the detector edges.

\begin{figure*}
  \includegraphics[width=\textwidth]{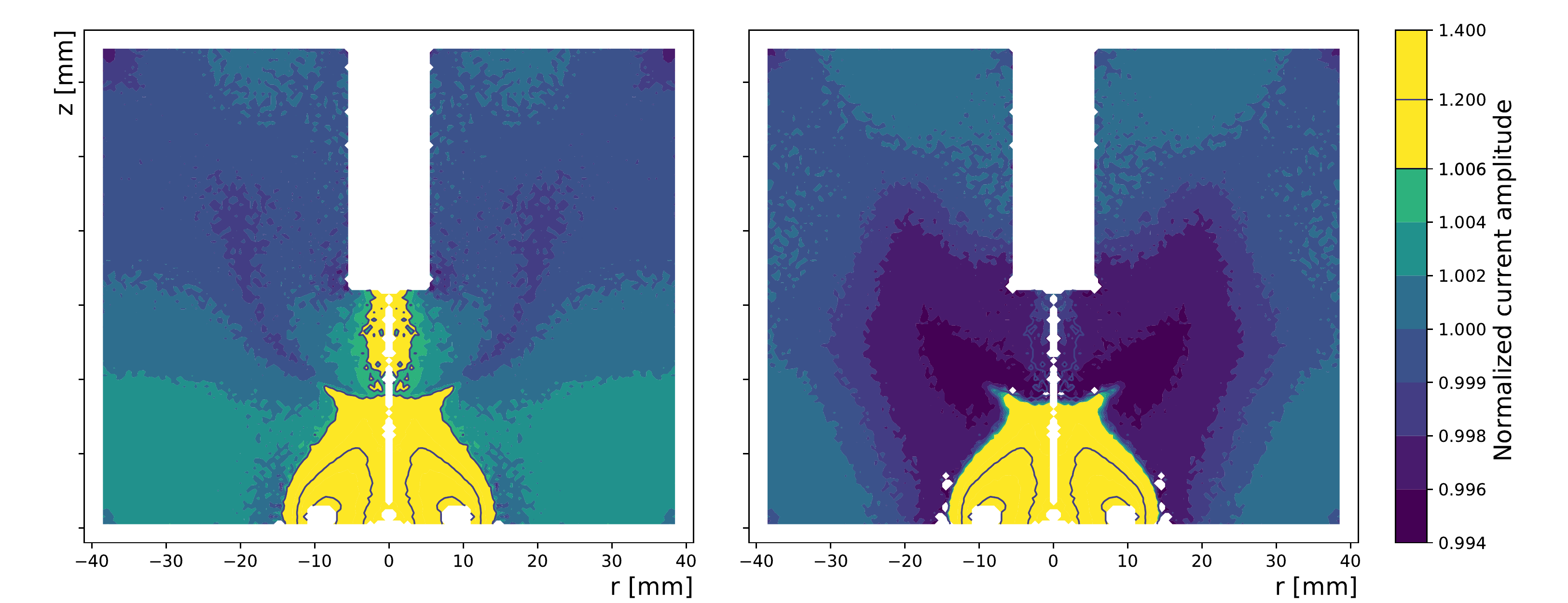}
\caption{$A/E$ maps from Monte Carlo \onbb~events. The left plot shows the values of $A/E$ normalized according to the energy correction (see \ref{app:A/E}) and the right plot shows the values after rise time correction.}
\label{fig:AoE_w-wo_corr}       
\end{figure*}

Motivated by the correlation shown in Fig. \ref{fig:RTvsAoE}, we explored the impact of a  first order linear correction of the $A/E$ value based on the rise time for each event. The $A/E$ maps before and after such correction are shown in Fig. \ref{fig:AoE_w-wo_corr}. The linear correction reduces the difference among A/E values: the volume that exhibits an $A/E$ value of $(1.000 \pm 0.002)$ increases from 71\% before correction to 89\% after. At the same time, it creates a bulk volume where $A/E$ values get lowered by almost 0.5\%. This is due to the interplay between collective effects, which combine in such a way that the cluster deformation (hence $A/E$) is not univocally associated to the length of the drift paths. 
In order to determine whether it is convenient to apply the rise time correction or not, we tested it on the simulations of $^{208}$Tl and \onbb. The results are reported in the second column of Tab. \ref{tab:SP}. The survival fraction of \onbb~events decreases after rise time correction from a value of $(86.1\pm 0.1$)\% to $(85.5\pm 0.1$)\%. 
In terms of background, the rise time correction increases the survival fraction of events at Q$_{\beta\beta}$ by $(1.5\pm0.3)$\%. 
The correction does not improve the overall efficiencies, but reduces the volume dependence of the PSA performance, possibly reducing the systematic uncertainties of the experiment. It might become more and more relevant as the detector volume keeps on increasing.

The  distribution of the $A/E$ and rise time from experimental data is shown in the coloured filled contour of Fig. \ref{subfig:RTvsA/E_DEP}, in comparison with simulations, represented by the gray contour lines.
The 0.3\% displacement in $A/E$ between the two blobs is well reproduced by our work. This is not the case if collective effects are not included. The excess in data at low values of $A/E$ is expected, as DEP events cluster on corners, where a fraction of events occurs in a transition layer where there is no electric field and the charge carriers move because of diffusion. This effect is not included in our simulation. The rise time is systematically underestimated by $\sim30$ ns in our simulation. This disagreement does not affect the conclusions of our work and could in principle be improved by tuning the unknown parameters of the crystal, such as the impurity profile along the symmetry axis, or the hole mobility. 

Experimental data for $^{208}$Tl have been collected using a $^{228}$Th source (a progenitor of $^{208}$Tl, details in \ref{app:Th}) and used to extract the survival fractions of the different classes of events, both before and after rise time correction. The numbers, reported in Tab. \ref{tab:SP}, show an agreement $<0.5\%$ with simulations for SEP and FEP events. Some tension appears when comparing the survival fractions of the Compton continuum at Q$_{\beta\beta}$. This can been traced back to inaccuracies in the positioning of the source. The distance between radioactive source and detector changes the fraction of multiple-site events from cascade of gammas (this was also observed in \cite{Agostini_SignalModeling}). This does not affect the populations of SEP and FEP events, since for them a statistical subtraction of the side-bands is performed (details in \ref{app:A/E}). The impact of the rise time correction on data, even if not statistically significant, reflects what is found with simulations, namely that it increases the acceptance of FEP and SEP events, as well as of background at Q$_{\beta\beta}$. In summary, the modeling developed reproduces the $A/E$ results within 0.2\% and hence its systematic uncertainties are lower than the impact of the collective effects that we wanted to study.

\section{Conclusions and discussion}
\label{sec:conclusion}

In this paper we discussed the collective effects in clusters of charge carriers in germanium detectors and the impact of such effects on signal formation, with particular focus on the consequences for \onbb~experiments with $^{76}$Ge. We determined that the deformation of the signal due to collective effects is relevant for detectors with long drift paths. In particular, we observed in the inverted coaxial geometry a position dependence of the standard pulse shape discrimination parameter used in \onbb~experiments ($A/E$). With the combined use of Monte Carlo and pulse shape simulations of $^{208}$Tl and \onbb s of $^{76}$Ge, we determined that such volume dependence does not impact the pulse shape discrimination performances significantly. This proved to be the case both using the standard $A/E$ analysis, and implementing a correction based on the reconstruction of the drift path.

\begin{figure}
  \includegraphics[width=0.5\textwidth]{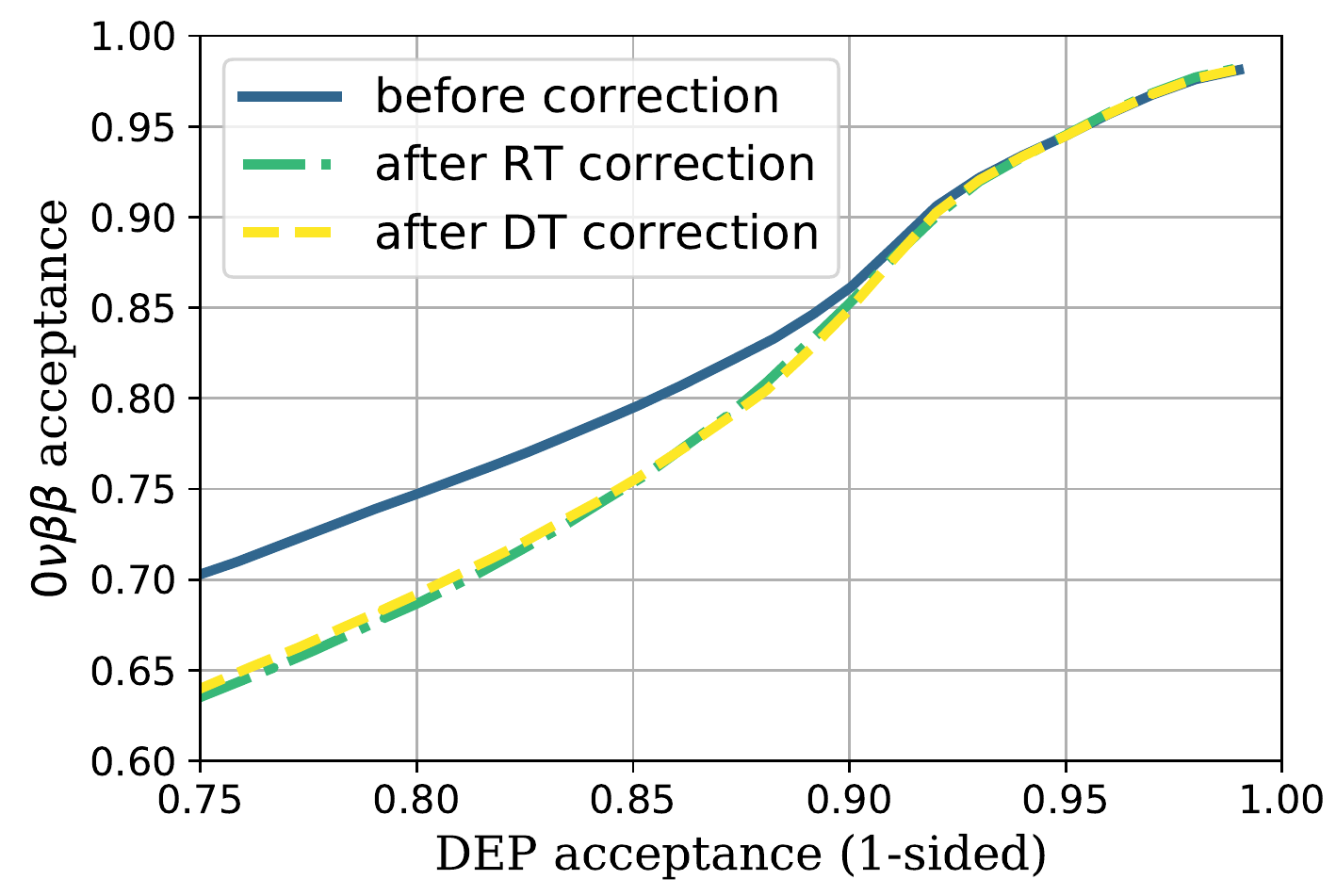}
\caption{Acceptance of \onbb~events as a function of DEP's, in the case of no-correction on $A/E$ (blue curve), or after rise time (green curve) and drift time (yellow curve) correction.}
\label{fig:GWD6022_SPDEPvs0nbb}       
\end{figure}

As detector volumes keep on increasing, the impact of collective effects on $A/E$ might become stronger~\cite{comellatoPisa}. Moreover, the background composition at Q$_{\beta\beta}$ will change, too, for different detector geometries. With such conditions, it is meaningful to compare detector performances at the same \onbb~acceptance. This could be used in the future to fix the $A/E$ cut on DEP events. A visual representation of the \onbb~acceptance as a function of the acceptance of DEP events is displayed in Fig.~\ref{fig:GWD6022_SPDEPvs0nbb}, both before and after rise time correction. No appreciable difference was observed when the true drift time (extracted from the simulations) was used for the correction. 


As planned by \legend, inverted coaxial detectors will be deployed in environments which are more challenging than a vacuum cryostat and exhibit different electronics noise conditions. In this work we explored the impact of a factor 5 higher noise level on pulse shape discrimination performances. This yields (for a cut at 90\% DEP acceptance) an increase in the \onbb~acceptance of 3\%, but at the same time an increase of 5\% in the background events surviving the $A/E$ cut at Q$_{\beta\beta}$. This is compatible with values of other BEGe detectors already in use in \gerda~\cite{Gerda:science}. We also explored the performances of inverted coaxial detectors with lengths in the range $8-9$ cm and determined that the performances are still compatible with those presented here. This fact, together with the other results of this work, confirms the inverted coaxial detectors as a high-performance design for the search for neutrinoless $\beta\beta$ decay.



\begin{acknowledgements}
We are very grateful to David Radford who developed SigGen as an open source project. SigGen is the software that we used to model the HPGe detector signal and included already the modeling of collective effect that we used to study the performance of our three detector geometries. We are also thankful to D. Radford for many suggestions and enlightening discussions during the work as well as his help during the preparation of this manuscript. We are also thankful to all the members of the GERDA and LEGEND collaborations for their valuable feedback.  This work has been supported in part by the European Research Council (ERC) under the European Union’s Horizon 2020 research and innovation programme (Grant agreement No. 786430 - GemX) and by the SFB1258 funded by the Deutsche Forschungsgemeinschaft (DFG).
\end{acknowledgements}

\appendix
\section{Details on simulations}
\label{app:sim}

This section deals with the technical details of the simulations carried out for this work. The physics model for \onbb~and $^{208}$Tl decays has been simulated within the \textsc{MaGe} software framework \cite{MaGe}, while the generation of signals in germanium detectors has been simulated using the SigGen software \cite{Radware}.

\subsection{Monte Carlo simulations}
\label{sec:mage}

The Monte Carlo simulations have been performed using the \textsc{MaGe} software, a \textsc{Geant}4 based framework oriented to low background experiments. \textsc{MaGe} gives the opportunity to select the track precision of the simulated particles, by choosing what is called the \emph{realm}. For this work, we used the \emph{DarkMatter} realm, in which the precision for gamma rays and e$^\pm$ are 5 $\mu$m and 0.5 $\mu$m, respectively. For a germanium detector, this means that every energy deposition of $\mathcal{O}(1)$ keV is stored as a hit. 

In order to estimate the signal acceptance and background rejection of our detector, we simulated \onbb s homogeneously distributed in the detector volume, and sources of $^{208}$Tl and $^{214}$Bi decaying at a distance of 20 cm from the detector. The location of the sources has been chosen to reduce the probability of detecting multiple gammas from the decay cascade, as this would result in an additional population of highly multiple-site events.

As in the experimental configuration, the DEP from $^{208}$Tl has been used to set the acceptance of single-site events. On the other hand, the samples of events @$Q_{\beta\beta}=2039$ keV from both $^{208}$Tl and $^{214}$Bi, plus the events from the Full Energy Peak (FEP) @2614.5 keV and Single Escape Peak (SEP) @2103.5 keV, are used as a background reference sample. In addition to the possibilities of an experimental setup, having a sample of simulated \onbb s allows to estimate the probability of accepting the sought-after signal. 

Also, from the Monte Carlo simulation of $^{208}$Tl decays, the energy dependence of the starting size of the charge carriers' cluster has been extracted. This has been done by means of the R90 parameter, which is defined as the minimum radius of the sphere which contains 90\% of the energy depositions. We selected 30 energy windows in the range $[1.0, 2.2]$ MeV, extracted the associated R90 value, and fitted the resulting energy dependence with a first order polynomial. The fitting function was then given as an input to SigGen, so that any energy is associated to an initial cluster size.

Finally, we simulated a collimated $^{241}$Am source shining on the side of the detector at different heights. The low energy (59.5 keV) gammas from it have been selected as samples of known and localized interaction position. The comparison of this dataset with analogous experimental data has been used to tune the physics parameters of the detector in SigGen.   

\subsection{Pulse Shape Simulations}
\label{sec:siggen}

SigGen is a software tool to simulate signals from germanium detectors. The signal generation consists in two parts: the first one, called \emph{fieldgen}, calculates the electric and weighting field of a given geometrical configuration. The second part, \emph{siggen}, simulates the signals generated by the drifting charges in the detector field.

For this work, the fields from \emph{fieldgen} are simulated on a 0.1 mm grid, and the signals fom \emph{siggen} are generated on a time step of 0.1 ns. 
The initial cluster size is chosen according to the information extracted from the Monte Carlo dataset of $^{208}$Tl (through the R90 parameter described in \ref{sec:mage}) and the crystal properties, such as the temperature and the impurity profile, are tuned using the combination of Monte Carlo and experimental data with a $^{241}$Am source.

The output of MaGe is a list of hits which constitute an event. In order to build the event-waveform (e.g. a \onbb~waveform), we generate signals for every hit and sum them all up, each with a weight corresponding to the energy deposited in the hit. The waveform obtained in this way, however, does not yet include collective effects, as very hit is processed separately. In order to take them into account, two steps more are needed. The first is to use the position of the first energy deposition to calculate the associated time spread of the cluster, $\sigma_\tau$. The second step is to convolute the event-waveform with a gaussian function of with $\sigma_\tau$.

Before the analysis, every waveform goes through the electronics response function developed by \cite{phd:panas}, whose parameters are again tuned using the combination of Monte Carlo and experimental data with a $^{241}$Am source. Furthermore, electronics noise, taken from our experimental setup, is added on top of the electronics processed waveform.

The relevant parameters for the analysis are calculated from differently processed waveforms. The rise time is extracted directly from the waveforms with noise, while the $A$ parameter has been calculated after applying 5 times a moving window average of 100 ns width. Finally, the energy $E$, given by Monte Carlo, has been smeared using a gaussian function whose width $\sigma_E$ has been inferred from the experimental resolution curve. 

For a comparison with data, the standard \gerda~analysis \cite{gelatio}, as described in \ref{app:A/E}, has been carried out for both simulated and experimental data. 

\section{A/E cut calibration}
\label{app:A/E}

This section describes in more detail the calibration procedure to set the $A/E$ cut and to calculate the survival fractions of different classes of event. This is entirely based on the works in \cite{DusanBEGe, phd:wagner}.

\subsection{The $^{228}$Th source}
\label{app:Th}

$^{228}$Th is the reference source in \onbb~experiments for multiple reasons. First, its daughter $^{208}$Tl has a gamma at 2.6 MeV which can undergo pair production in the interaction with the detector. When this is the case, the two 511 keV photons from the annihilation of the positron can either be absorbed in or escape the detection volume. In case both are absorbed in the detector, their energies sum up to that of the electron, thus falling into the Full Energy Peak (FEP) at 2.6 MeV. When one of the two escapes detection, the detector measures 2.6 - 0.511 MeV and the event is referred to as Single Escape Peak (SEP). If pair creation occurs on corners, there is a significant probability that neither of the 511 keV photons deposit any energy in the detector. This case is known as Double Escape Peak (DEP) and is of particular importance for \onbb~experiments, as it consists of an electron and positron depositing 1.592 MeV in the detector, thus resembling the physics of the energy deposition from \onbb~(of course, with different energy and geometrical distribution). For this reason, DEP events are used as a proxy for signal-like events.

On the other hand, SEP events, being composed of an electron-positron pair and a gamma, are characterized by two (normally) spatially separated energy depositions. Those events, together with those from the FEP of $^{208}$Tl and $^{212}$Bi, which are mainly composed of multiple Compton scattering, are used as reference to estimate the event discrimination performance of a detector.

Furthermore, what makes $^{228}$Th also a valuable source for \onbb~search, is that at the energy of $Q_{\beta\beta}= (2039\pm35)$ keV, the spectrum is composed of events with different topologies, among which, a fraction can undergo single Compton scattering, and thus mimic the signal of a \onbb. This is an irreducible background for pulse shape discrimination alone, but it is mitigated in \gerda~by active veto systems which tag energy depositions outside the detector volume \cite{Gerda:science}. 

\subsection{$A/E$ Analysis}
\label{sec:SSE}

The $A/E$ analysis is based on a single parameter that is the maximum value of the current signal ($A$), normalized by the total deposited energy ($E$). In case of a single energy deposition, $A/E$ exhibits a value which is higher than the case of a multiple energy deposition. This is due to the fact that a multiple energy deposition distributes the total charge in several clusters, each generating a current proportional to the charge contained in the cluster. 

As the starting size of the cluster increases with energy, its time spread (our $\sigma_\tau$ parameter) gets larger, generating lower values of $A/E$. This energy dependence is estimated by selecting an arbitrary number of energy regions in the Compton continuum in the range $[1.0, 2.3]$ MeV and extracting the $A/E$ values for each region. The dependence on energy is then fitted and corrected with a linear interpolation.

The standard analysis uses DEP events from $^{208}$Tl as a proxy of single energy depositions and fixes a low cut value for $A/E$ by setting the acceptance of DEP events to 90\%. With this value, it computes the number of events surviving the cut for different event classes: this is referred to as a 1-sided cut. In addition, in order to reject surface events from regions which are close to the p$^+$ electrode (which are potentially coming from surface contamination), it computes the mean $\mu$ and width $\sigma$ of the distribution of $A/E$ and sets the high cut to the value of $\mu + 4\sigma$: this procedure is referred to as a 2-sided cut. 

In order to extract the correct survival fractions of DEP, SEP and FEP events, the standard analysis gets rid of the Compton scattering events which lie in the same energy region of interest by a statistical subtraction: for every peak, two sidebands are selected (at lower and higher energy), whose $A/E$ distribution is then subtracted from that of the peak of interest.


\printbibliography[heading=bibintoc]

\typeout{get arXiv to do 4 passes: Label(s) may have changed. Rerun}

\end{document}